\documentclass[preprint]{aastex}
\usepackage{psfig}
\slugcomment{Draft: \today}
\shorttitle{Isolated Galaxies in SDSS DR1}
\shortauthors{Allam et al.}
\begin{document}
\title{A Catalog of Very Isolated Galaxies from the SDSS Data Release~1}
\author{
Sahar S. Allam\altaffilmark{1,2},
Douglas L. Tucker\altaffilmark{2},
Brian C. Lee\altaffilmark{3},
J.\ Allyn Smith\altaffilmark{4,5}
}
\altaffiltext{1}{New Mexico State University, Department of Astronomy, 
P.O. Box 30001, Dept 4500, Las Cruces, NM 88003}
\altaffiltext{2}{Fermi National Accelerator Laboratory, P.O. Box 500, 
Batavia, IL 60510}
\altaffiltext{3}{Lawrence Berkeley National Laboratory, 1 Cyclotron Rd, 
Berkeley CA 94720-8160}
\altaffiltext{4}{Los Alamos National Laboratory, NIS-4, MS D448, 
Los Alamos, NM 87544-1724}
\altaffiltext{5}{University of Wyoming, Dept.\ of Physics \& Astronomy, 
P.O.Box 3905, Laramie, WY 82071}
\begin{abstract}
We present  a  new catalog of   isolated galaxies obtained through  an
automated systematic search.  These  2980 isolated galaxies were found
in $\approx$  2099 deg$^{2}$ of  sky in  the Sloan  Digital Sky Survey
Data  Release  1  (SDSS  DR1) photometry.   The  selection  algorithm,
implementing a variation on the criteria developed by Karachentseva in
1973, proved to be   very efficient and   fast.  This catalog will  be
useful for  studies of the   general galaxy characteristics.   Here we
report on our results.
\end{abstract}

\section{Introduction} 

Over  the past  few decades,  observational  and theoretical work  has
shown that  truly isolated ``field'' galaxies,  if they exist  at all,
are a rarity in the universe, comprising less than 5\% of all galaxies
(\citealt{Adams80}).  Rare  though  they  may  be, they  serve as   an
important  comparison sample in studies of  the effects of environment
on   galaxy    morphologies   and    star  formation     rates  (e.g.,
\citealt{Adams80,Haynes80,Haynes84,Koopmann98}).      Truly   isolated
galaxies, which may have experienced no major interactions in billions
of years, can act as a zeropoint in these studies.

Furthermore, isolated  galaxies are  interesting in their   own right.
Recent studies of   isolated galaxies include those by  \citet{Aars02,
Aars03}, who looked at the photometric and spectroscopic properties of
extremely isolated  elliptical galaxies in the \citet{Karachentseva73}
catalog;  \citet{Pisano02}  and   {\citet{Pisano03}, who performed  an
\ion{H}{1} survey for the  gaseous  remnants of the  galaxy  formation
process around  nearby ($\la$  30$h^{-1}$~Mpc) isolated galaxies  they
identified  in  the  \citet{Tully88}'s  ``Nearby  Galaxies  Catalog'';
\citet{Sauty03}, who measured  the molecular gas  mass of 99  isolated
late-type galaxies  in  the   \citet{Karachentseva73}   catalog  using
observations   of  the   $^{12}$CO(1-0)  line;  \citet{Stocke04},  who
measured  the the luminosity functions  of  isolated elliptical and S0
galaxies in the \citet{Karachentseva73} catalog; and \citet{Varela04},
who studied  the properties of disk galaxies  in a catalog of isolated
galaxies  that they extracted from  a volume-limited sample of the CfA
Redshift Survey \citep{Huchra00}.

A  simple  method  of  identifying   isolated  galaxy  candidates  was
described by \citet{Karachentseva73}.    She selected all  galaxies in
the   {\em  Catalog    of    Galaxies  \&    Clusters  of  Galaxies\/}
(\citealt{Zwicky68}; Zwicky  Catalog) whose nearest neighbors, of size
within a factor of four of the same  diameter, lie further than twenty
diameters  away.  Then she compiled  her catalog from work with prints
of the Palomar    Sky  Survey.  The original  catalog   contains  1052
candidate isolated galaxies,  which was later  reduced to 893 galaxies
(\citealt{Karachentseva80}).   She   also showed  that  a  significant
number of   the final catalog  members  are located in regions  of low
density in the periphery of superclusters.  (Note that
\citealt{Leon03} have  recently updated  the \citealt{Karachentseva73}
catalog with improved galaxy positions.)

\citet{TurnerGott75} suggested    an    interesting  prescription  for
isolating the classical uniform  field  population of galaxies in  the
Zwicky catalog.  They placed  galaxies brighter than 14th magnitude in
two  classes,  the  associated (A)  that  have  at least one  neighbor
brighter than $m=14$ within 45$\arcmin$, and the single (S) which have
none.    The  S   galaxies, 43\%   of   the   sample, appear uniformly
distributed across the sky, as expected for field galaxies.  Later
\citet{HuchraThuan77}  determined   that only  12 out of 1088 S galaxies
($\approx$1\%) of the Turner \& Gott sample can be considered as truly
isolated.  Turner \& Gott S  galaxies appeared single because the 14th
magnitude cutoff misses fainter companions  and the angular separation
criterion of $45\arcmin$ mistakes members of nearby groups of galaxies
with large angular extent for singles.

These earlier efforts for  identifying homogeneous samples of isolated
galaxies   were magnitude limited to  $B<15$.    The Sloan Digital Sky
Survey (SDSS; \citealt{York00}), which will eventually cover up to one
quarter of  the  sky with uniform photometry  in  five filters down to
$g\approx 22$, makes  for  an  obvious  hunting ground  for   isolated
galaxies to much greater depths and volumes.  Using a modified version
of Karachentseva's  isolation   criteria, we  have  extracted  such an
objectively defined catalog of  isolated galaxies from the photometric
catalog of the SDSS Data Release 1 (DR1;  Abazajian et al.\ 2003).  We
present this catalog as follows.  In \S~\ref{sec:DATA} we describe the
region  of  the   sky   used  for  this    preliminary    search.   In
\S~\ref{sec:SelectionCriteria}, we   describe the catalog construction
techniques,  and we present  the  catalog in \S~\ref{sec:catalog}.  In
\S~\ref{sec:conclusions} we conclude and describe our future plans.

\section{The Data}
\label{sec:DATA}

The SDSS is a digital photometric and spectroscopic survey which will,
when  completed, cover one $\pi$ steradian  of the celestial sphere in
the North Galactic Hemisphere and  an additional 225 square degrees in
the South    Galactic  Hemisphere.   The  photometric   mosaic  camera
(\citealt{Gunn98}) images  the sky by  scanning along great circles at
the sidereal rate.  The imaging   data are produced simultaneously  in
five  photometric bands ($u$, $g$, $r$,  $i$,  and $z$, with effective
wavelengths   bands of  $\lambda$4770,  $\lambda$6231,  $\lambda$7625,
$\lambda$9134 $\AA$  respectively,    cf.\  see \citealt{Fukugita96}),
under  photometric conditions   (\citealt{Hogg01})  and  is  targeting
10$^{6}$ objects for spectroscopy (Blanton et al.\ 2003) most of which
are galaxies with $r$ band magnitude $<$ 17.77 (\citealt{Strauss02}).

The SDSS data   are  reduced  by  highly  automated  photometric   and
spectroscopic  reduction  pipelines (see  \citealt{Stoughton02}).  The
astrometric calibration is  automatically performed by a pipeline that
obtains  absolute     positions   to     better   than   0.1    arcsec
(\citealt{Pier03}),     sources    are    identified,   deblended  and
photometrically measured (\citealt{Lupton02}), and then the magnitudes
are calibrated   to a standard star network   approximately  in the AB
system   (\citealt{Smith02}).    After  selecting    the targets   for
spectroscopy     (\citealt{Eisenstein01};         \citealt{Strauss02};
\citealt{Richards02}),       spectroscopic      fibers    are   placed
(\citealt{Blanton03}),    and    spectroscopic   data  reduction   are
automatically performed to measure redshift.

On 2003 April 17, the SDSS team made public their Data Release 1 (DR1;
\citealt{Abazajian03}) to the astronomical community.  SDSS DR1 covers
$\approx$ 2099~deg$^{2}$   of five-band imaging  data   of the Sky and
includes spectra, with   derived spectroscopic parameters,   for 22108
stars, 133996  galaxies and 18678  quasars.  For more details  see the
on-line documentation at http://www.sdss.org/dr1.

We used  all  objects  classified  as  galaxies  by  the SDSS  imaging
reduction software  ({\tt  photo}, \citealt{Lupton02}) from  the  SDSS
public database   as our base   catalog from  whence we  extracted our
catalog of isolated galaxies.
%
\section{Selection Criteria}
\label{sec:SelectionCriteria}
We have developed a  systematic search criterion for isolated galaxies
in SDSS DR1.  We have made use of a computer code embodying a slightly
modified  version  of Karachentseva's  (1973)  criteria.  Under  these
criteria,  a galaxy  $i$  with angular  diameter  $a_i$ is  considered
isolated if the projected sky separation $x_{i,j}$ between this galaxy
and any neighboring galaxy $j$ of angular diameter $a_j$ satisfies the
following two relations:
\begin{eqnarray} 
x_{i,j} \ge  20 \times a_j
\end{eqnarray} 
\vspace{-.65cm}
\begin{eqnarray} 
\frac{1}{4} a_j \leq a_i \leq  4 \times a_j 
\end{eqnarray}

As noted above, we selected for  our base catalog all SDSS DR1 objects
which  were classified  as galaxies  by  {\tt photo};  we imposed  the
addition  requirements  that  these  objects have  $g$-band  Petrosian
magnitudes $g  > 0$ and Petrosian  radii $R > 0$.   Under our modified
Karachentseva criteria,  a galaxy $i$ with a  $g$-band magnitude $g_i$
and $g$-band  Petrosian radius $R_i$  is considered to be  isolated if
the projected  sky separation between this galaxy  and any neighboring
galaxy $j$ satisfies:
\begin{eqnarray} 
x_{i,j} \ge 40 \times R_j
\end{eqnarray}
\vspace{-.65cm}
\begin{eqnarray} 
| g_i - g_j | > 3.0 
\end{eqnarray}
Note  that a magnitude  difference of  3 is  about a  factor of  16 in
brightness;  thus, equation~4  roughly approximates  equation~2  for a
galaxy with a flat surface-brightness profile.

We considered only candidate isolated  galaxies in $g$-band  magnitude
limit  (after correcting for     Galactic extinction using   the  dust
distribution estimated  by  \citealt{Schlegel98}) of $16.0\leq g_i\leq
21.0$.  (As we will  see  in \S~\ref{sec:completness},  our  selection
criteria effectively reduce this magnitude range to $16.0 \leq g_i \la
19$.)   Using  these  modified  criteria, we found   a   total of 3813
candidates in 2099~deg$^{2}$.  To remove spurious  objects due to poor
image  de-blending, one of  us (SSA)  inspected  all candidate by eye.
She  also used {\tt   SExtractor} (\citealt{BertinArnouts96})  on  the
$g$--band SDSS FITS  images to double-check galaxy identification. 923
candidates were removed due to: bright stars misidentified as galaxies
(320), part   of  bright galaxy (50),  diffraction   spike from nearby
bright star  (417), and finally  136 were found  to be  diffuse light.
After all rejections and verifications,  the final number of candidate
isolated galaxies  left for inclusion   in this catalog was  2980,  or
$\approx$ 1.4 per sq~deg.

We  note that  \citet{Prada03} have  also  extracted isolated galaxies
from  the   SDSS.   In   their  case,  however, they    used  the SDSS
spectroscopic  sample of   galaxies,  which  is  restricted mostly  to
magnitudes $r  <   17.77$.  Using  the velocities of   small satellite
galaxies  surrounding the isolated   SDSS galaxies, they were able  to
measure the dark matter  profile for relatively  unperturbed galaxies.
Our sample is complementary to theirs, in that our sample goes fainter
and that our  isolation criteria is somewhat  more restrictive.  Also,
as our criteria closely  mimic  those of \citet{Karachentseva73},  our
sample    has a close connection  with    previous studies of isolated
galaxies.

\section{Catalog}\label{sec:catalog}
In  Table~\ref{tab1}  we  list the general    properties  of the  2980
isolated galaxies: Col.\ (1) a   running identification number,  Col.\
(2) galaxy name (following the IAU-designated SDSS naming convention),
Col.\ (3)  the $g$--band Petrosian   magnitude corrected for  Galactic
extinction, Col.\ (4) the reddening in the  $g$ band as estimated from
the \citet{Schlegel98} reddening    maps,  Col.\ (5)  the    $g$--band
Petrosian radius, Col.\ (6) the  galaxy redshift (when available), and
Col.\ (7) the concentration index.  In Table  \ref{tab2}  we summarize
the mean and median properties  of the isolated  sample.  We also plot
the distribution  of   apparent  magnitudes,   apparent colors,    and
Petrosian radii in Figures \ref{figlabelhismag}, \ref{figlabelhisCol},
\& \ref{figlabelhisradius} respectively.

Figure \ref{figlabelcol} presents the (apparent) $u-g$ vs.  $g-r$, the
$g-r$ vs.  $r-i$, and the $r-i$ vs. $i-z$ color-color diagrams for all
galaxies  classified   as  isolated   by  our  criteria.    The  color
distributions compare  well with those  of Shimasaku et  al.\ (2001)'s
analysis  of bright  SDSS galaxies  (see their  Fig. 7)  and  those of
Yasuda  et al.   (2001)'s galaxy  number counts  analysis of  the SDSS
(their Fig. 2).

In  Figure  \ref{figlabelpolar}  we  show a   polar  view of  the 1886
isolated sample  with   spectroscopic  redshift  (red dots)   and  for
comparison   plotted  all   the  SDSS   DR1    galaxies with  redshift
information.   The   mean   redshift   for    our sample  is   $z_{\rm
mean}$=0.0642, which corresponds  to a comoving  distance of $\approx$
190$h^{-1}$ Mpc, and  the maximum redshift $z_{\rm max}$=0.2374, which
corresponds  to  a  a comoving   distance  of 669$h^{-1}$   Mpc for an
$(\Omega_{\rm M}=0.3, \Omega_{\Lambda}=0.7)$ cosmology.

We calculated the absolute magnitudes and colors for the 1886 isolated
galaxies with spectroscopic redshifts  by assuming a flat cosmological
model with $\Omega_{\rm M} = 0.3$, $\Omega_{\Lambda}  = 0.7$, and $H_0
= 100h$~km~s$^{-1}$~Mpc$^{-1}$   and by applying k-corrections  to the
de-reddened galaxy magnitudes by means of  the publicly available {\tt
kcorrect (v1.10)}  code  of  \citet{Blanton03}, where  the  luminosity
distances were estimated using the analytical relation of
\citet{Pen99}.   We  plotted  the  absolute $M_{g}-M_{r}$  color  vs.\
spectroscopic  redshift for  the  galaxies (Fig.~\ref{figlabelzMgMr}).
In Figures~\ref{figlabelMg}, \ref{figlabelMr}, and \ref{figlabelMall},
we plot rest-frame color-magnitude  and color-color diagrams for these
isolated galaxies.

\subsection{Concentration Index}
\label{sec:CI}
The concentration index  (CI)  is defined by   the  ratio of the   two
$r$--band Petrosian radii, $CI \equiv  r90/r50$, where $r90$ and $r50$
correspond  to the radii  at which the integrated  fluxes are equal to
90\%   and   50\%  of the     $r$--band Petrosian flux,  respectively.
\citet{Shimasaku01}  report  that   this  $CI$   parameter shows   the
strongest correlation with visually-classified morphology among simple
photometrically-defined parameters.  Spiral galaxies are usually found
to  have small $CI$ ($\leq$  2.5) whereas ellipticals have higher $CI$
($>$2.5).    We thus separate morphologies  into  early and late types
according as  $CI\leq2.5$    or $CI>2.5$,  which  corresponds  to  the
division at S0/a.  In Fig \ref{figlabeCi}  we show the distribution of
concentration  indices for isolated galaxies.  We  find  our sample of
isolated galaxies to contain 1414$\pm$38  (47\%) late type Spirals and
1566$\pm$40 (53\%)  early  type galaxies.   (Error bars   are based on
$N^{1/2}$ statistics.)  Note  that late--type Spirals  only marginally
outnumber early--type galaxies in our sample.

\subsection{The Atlas}
\label{sec:Atlas}
Due to the large number of galaxies in  our catalog, we do not provide
a   hardcopy atlas.  Instead, we have   prepared an online color atlas
from   the   SDSS  DR1   located   at  our   public  URL. \footnote{{\tt
http://home.fnal.gov/$\sim$sallam/ISOLATED}}

\subsection{Completeness}
\label{sec:completness}
As is typical for isolated galaxy catalogs based upon observed angular
sizes and  distances between  galaxies, our catalog  --- like  that of
\citet{Karachentseva73}  ---   is  more  representative   than  it  is
complete, due to the unintended  exclusion of galaxies that would have
otherwise    met   the    isolation   criteria    but    contained   a
foreground/background    galaxy    within    its   isolation    radius
\citep{Karachentseva80,Sauty03,Stocke04}.  

We      can  look  at   this      in   two   ways.   First,   consider
Figure~\ref{figcomplet}.  Here, we plot the number counts for all SDSS
DR1 galaxies and the  number counts for just  the galaxies in our SDSS
DR1 isolated galaxy catalog.  In both cases, the number counts are for
0.1  mag bins and  the  error bars associated    with each symbol  are
Poission ($\sqrt{N}$).   The number counts  for the sample of all SDSS
DR1 galaxies exhibit a linear behavior in $\log N$ vs. $g$, showing no
obvious signs of evolution in  this magnitude range ($g=16-21$).   The
number counts for  the isolated  sample,  however, clearly reaches   a
maximum  around $g=17$ and  drops off essentially to  zero  for $g \ga
19$.  If  we assume that the fraction  of  true isolated galaxies does
not  evolve substantially over this  magnitude range, it is clear that
fainter isolated galaxies are missing from our sample.

Second, to see that this  incompleteness is inherent in the  selection
criteria,  consider   Figure~\ref{figCriterionTest}.  Here,   we  have
plotted  the minimum  value  of the scaled separation  $x_{i,j}/R_{j}$
(see  eq.~3) for each of 3354  galaxies in a   subset of the SDSS DR1.
(For  each of these 3354 galaxies,  only neighbors  within 3.0~mag are
considered; see eq.~4.)  Galaxies whose nearest neighbor (in the sense
of $x_{i,j}/R_{j}$)  lies more than  $40 \times R_{j}$ away  --- i.e.,
above the  horizontal dashed line in Figure~\ref{figCriterionTest} ---
would  be considered isolated.   Note  that the $x_{i,j}/R_{j} \ge 40$
isolation    criterion    is quite   restrictive.   Essentially,  only
``outliers''  meet it,  and  the  number  of  outliers decreases  with
increasing $g$  magnitude; in the  DR1 subsample plotted,  no galaxies
fainter than $g\approx 18.5$ meet this criterion.

\subsection{Nearest Neighbor and Comparison with a Field Sample}
\label{sec:NNComparison}
To test our isolation criteria  for our isolated galaxies, we used all
SDSS  DR1 galaxies  with redshifts  to  construct the  search for  the
nearest-neighbor distance, $d_{min}$,  which represents the separation
in  the 3-dimensional  redshift  space.  To  calculate the  separation
between each isolated galaxy  and its nearest neighboring SDSS galaxy,
we introduce the concept of  a distance field (Stavrev 1990).  Such an
approach has been applied also by \citet{Frisch95}, \citet{Lindner95},
and by \citet{AikioMaehoenen98}.

Let $ISO$ be an  isolated galaxy with Cartesian coordinates $x_{ISO}$,
$y_{ISO}$, $z_{ISO}$  in a 3-D coordinate  system, and let  $j$ be any
{\em other}  SDSS galaxy with Cartesian  coordinates $x_{j}$, $y_{j}$,
$z_{j}$.  To speed the calculation, we only consider SDSS galaxies $j$
which  lie within  a one  square degree  box centered  on  an isolated
galaxy's RA  and DEC.   For each isolated  galaxy the distance  to its
nearest neighboring object is computed as:

\begin{equation}
d_{min} =  \mbox{min}[\sqrt{(x_{ISO}-x_{j})^{2}+(y_{ISO}-y_{j})^{2}+(z_{ISO}-z_{j})^{2}}].
\end{equation}

The mean  nearest-neighbor distance for the 1839  isolated galaxies is
4.18$\pm$0.24  $h^{-1}$ Mpc,  and the  median is  1.144  $h^{-1}$ Mpc.

In order to have a fair comparison, we have constructed a random field
sample taken from  the SDSS DR1 redshift sample  having the exact same
number of  galaxies and having  the same redshift distribution  as the
isolated  galaxies sample.   Figure  \ref{figlabelneighbor} shows  the
distribution  of the nearest  neighbor distances  for the  1839 (solid
line)  isolated  galaxies and  field  sample  (dash  line).  The  mean
nearest-neighbor  distance for  1839 field  galaxies  is 3.12$\pm$0.19
$h^{-1}$ Mpc, and the  median is 0.978$h^{-1}$ Mpc.  A one-dimensional
Kolmogorov-Smirnov (KS)  test (Press et al.\ 1992)  indicates that the
distributions of nearest neighbors for the isolated and field galaxies
have  only a  0.0003\%  probability  that they  derive  from the  same
parent (Fig.~\ref{figlabelneighborKS}).

It is not surprising that the field sample has smaller mean and median
nearest  neighbor distances, since  the  field  sample should be  more
clustered   on average than the  isolated  galaxies --  after all, the
field sample is likely contaminated at about the 10\% level by cluster
galaxies, whereas the typical isolated  galaxy will more likely sit in
a wall or filament or the outer parts of a cluster.

\section{Conclusions}
\label{sec:conclusions}
A key problem in astronomy involves the role of the environment in the
formation and evolution of galaxies. In  order to answer this question
it   is necessary to characterize  a  reference sample  with a minimum
influence from the environment  so  that its evolution is   completely
determined by nature.

A fundamental issue in  galactic evolution is the  relative importance
of  initial   conditions vs.\  environment.   To  address the  role of
non-cluster environments,   we  present a  new   catalog  of  isolated
galaxies in the  SDSS DR1 data. 

At a  detection rate of  1.4  isolated galaxies per square  degree, we
expect that  the   final catalog, based upon   the  a completed   SDSS
covering up  to one-quarter of the sky,  will contain on the  order of
$\approx$ 14,000 galaxies.   This catalog will allow statistical study
of the properties of the  Interstellar Medium as  a function of galaxy
environment,    and its  relation to  star   formation, morphology and
luminosities, as well as nuclear  activity frequency (e.g.   Lisenfeld
et al.  2003).  This catalog will also be useful for future studies of
dark matter   density profile  of  isolated  galaxies (e.g.   Prada et
al. 2003).   Finally, this  catalog  will offer a  sample  of galaxies
which can greatly   aid in the investigation  of  galaxy evolution and
galaxy formation.

\acknowledgments

This paper is dedicated  to the memory of Prof.  Gamal El Din.  During
his career at NRAIG, the late  Gamal El Din  has had a major impact on
the Astronomy research in Egypt.

Funding for the creation and distribution of the SDSS Archive has been
provided  by  the  Alfred   P.  Sloan  Foundation,  the  Participating
Institutions, the  National Aeronautics and  Space Administration, the
National  Science  Foundation,  the  U.S. Department  of  Energy,  the
Japanese Monbukagakusho, and the Max Planck Society. The SDSS Web site
is http://www.sdss.org/.

The SDSS is managed by the Astrophysical Research Consortium (ARC) for
the Participating Institutions. The Participating Institutions are The
University of Chicago, Fermilab, the Institute for Advanced Study, the
Japan Participation  Group, The  Johns Hopkins University,  Los Alamos
National  Laboratory, the  Max-Planck-Institute for  Astronomy (MPIA),
the  Max-Planck-Institute  for Astrophysics  (MPA),  New Mexico  State
University, University of Pittsburgh, Princeton University, the United
States Naval Observatory, and the University of Washington.

This  research has made   use of the NASA/IPAC Extragalactic  Database
(NED),  which  is operated by   the Jet Propulsion  Laboratory, at the
California Institute of Technology,  under contract with the  National
Aeronautics and Space Administration.


\begin{deluxetable}{l@{ }r@{ }c@{ }c@{ }c@{ }c@{ }c}
\tabletypesize{\tiny}
\tablecaption{The SDSS DR1 Isolated galaxies \label{tab1}}
\tablewidth{0pt}
\tablenum{1}
\tablehead{
 \colhead{ID} & 
 \colhead{SDSSID} & 
 \colhead{g} &
 \colhead{A(g)} & 
 \colhead{Radius g} & 
 \colhead{z} & 
\colhead{ci} \\
 \colhead{} & 
 \colhead{} & 
 \colhead{} &
 \colhead{} & 
 \colhead{arcsec} & 
 \colhead{} & 
\colhead{} \\
\colhead{(1)} &
\colhead{(2)} &
\colhead{(3)} &
\colhead{(4)} &
\colhead{(5)} &
\colhead{(6)} &
\colhead{(7)} \\
 }
\startdata
1     & SDSS J000015.73-085327.92 &      16.639  &      0.145  &      6.256  &  0.055887 &  2.861 \\ 
2     & SDSS J000044.14+004022.80 &      16.203  &      0.106  &      6.823  &  0.062586 &  2.654 \\ 
3     & SDSS J000047.12-104809.00 &      16.879  &      0.133  &      5.913  &  0.061544 &  2.258 \\ 
4     & SDSS J000052.08-103512.84 &      17.187  &      0.140  &      8.356  &  0.062488 &  2.336 \\ 
5     & SDSS J000104.91-090915.55 &      16.448  &      0.141  &      9.741  &  0.055826 &  2.608 \\ 
6     & SDSS J000123.74-005106.95 &      17.248  &      0.132  &      5.173  &  0.078192 &  3.128 \\ 
7     & SDSS J000135.73-002342.63 &      16.736  &      0.135  &      9.146  &  0.078164 &  2.283 \\ 
8     & SDSS J000245.38-103429.28 &      18.049  &      0.148  &      3.234  &  0.058669 &  2.505 \\
\enddata
\end{deluxetable}

\clearpage

\begin{deluxetable}{lrr}
\tablecaption{SDSS DR1 Isolated Galaxy Mean Properties. \label{tab2}}
\tablewidth{0pt}
\tablenum{2}
\tablehead{
\colhead{}  & 
\colhead{mean}  & 
\colhead{median}
}
\startdata
$u$ & 18.359$\pm$0.0146 &18.262\\
$g$ & 16.938$\pm$0.0115 &16.869\\
$r$ & 16.327$\pm$0.0129 &16.245\\
$i$ & 15.996$\pm$0.0133 &15.910\\
$z$ & 15.821$\pm$0.0144 &15.722\\
 &  &  \\
$u-g$ &1.393$\pm$0.007&1.346 \\
$g-r$ &0.613$\pm$0.005&0.634 \\
$r-i$ &0.331$\pm$0.003&0.352 \\
$i-z$ &0.176$\pm$0.004&0.211 \\
 &  &  \\
$R_{Petro}u$ &9.072$\pm$0.176& 6.746\\
$R_{Petro}g$ &7.933$\pm$0.062& 7.344 \\
$R_{Petro}r$ &7.811$\pm$0.069& 7.048 \\  
$R_{Petro}i$ &7.516$\pm$0.063& 6.878 \\   
$R_{Petro}z$ &7.359$\pm$0.073& 6.613 \\   
&  &  \\
$\rm {CI}$&   2.528$\pm$0.008 &  2.471  \\
&  &  \\
	redshift (1886)   &   0.064237$\pm$0.00071&   0.0615\\   
	$M_{g}$ (1886)	  & -19.418$\pm$0.0315&  -19.675 \\ 
	$M_{r}$ (1886)	  & -19.984$\pm$0.0322&  -20.251 \\    
$M_{u}$ - $M_{g}$ (1886)  &   1.249$\pm$0.0071&    1.228 \\ 
$M_{g}$ - $M_{r}$ (1886)  &   0.566$\pm$0.0040&    0.583 \\ 
$M_{r}$ - $M_{i}$ (1886)  &   0.276$\pm$0.0035&    0.294 \\ 
$M_{i}$ - $M_{z}$ (1886)  &   0.217$\pm$0.0038&    0.233 \\ 
\enddata
\end{deluxetable}
\begin{figure*}
\centering
\makebox[160mm]{\psfig{file=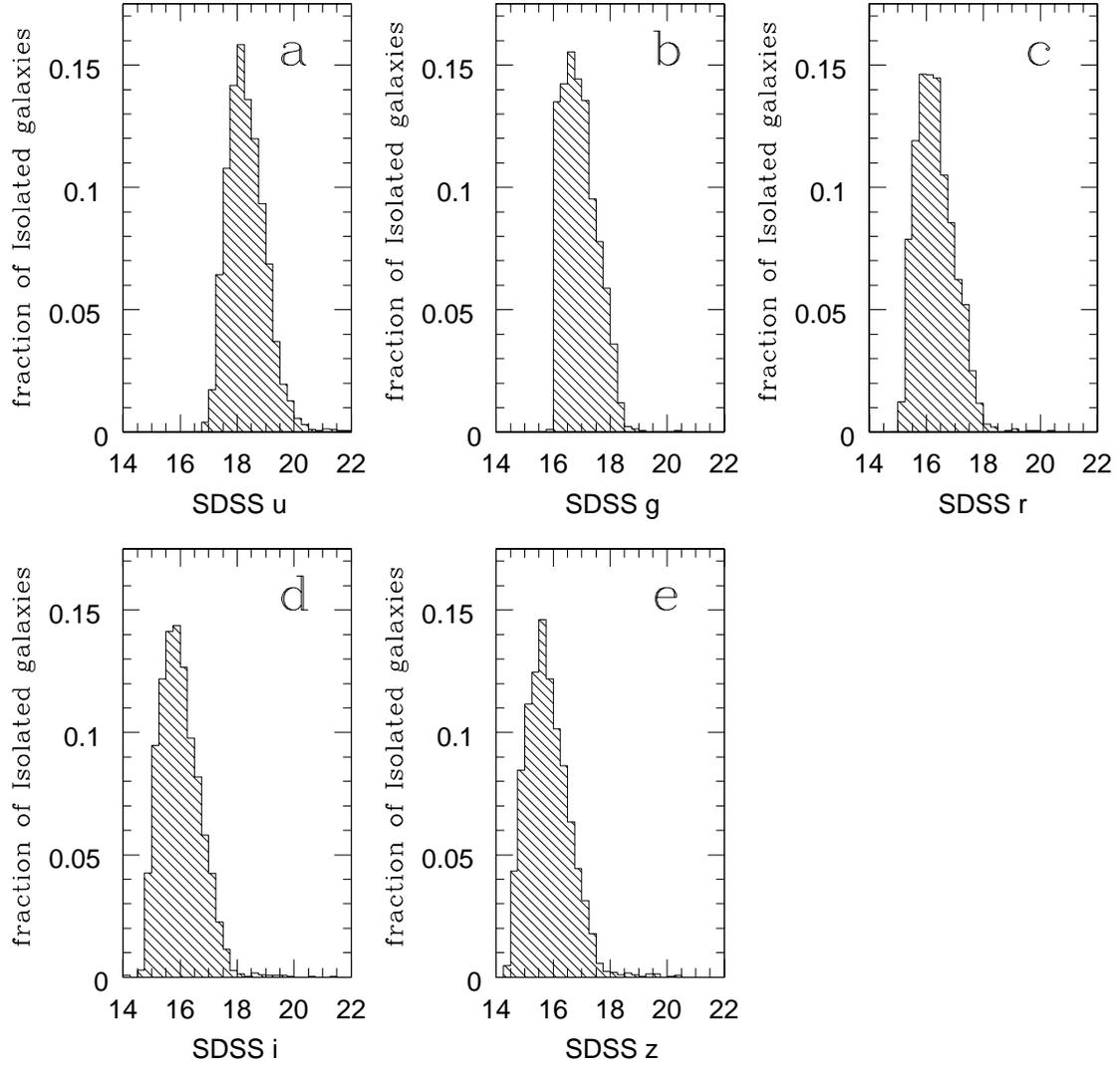,width=155mm,height=155mm,rwidth=150mm,clip=,angle=0,silent=}}
\caption{The distributions of the apparent magnitudes of the SDSS DR1 isolated galaxies sample in the 5 SDSS filters.  
Magnitudes are corrected for Galactic extinction.}\label{figlabelhismag}
\end{figure*}
\begin{figure*}
\centering
\makebox[160mm]{\psfig{file=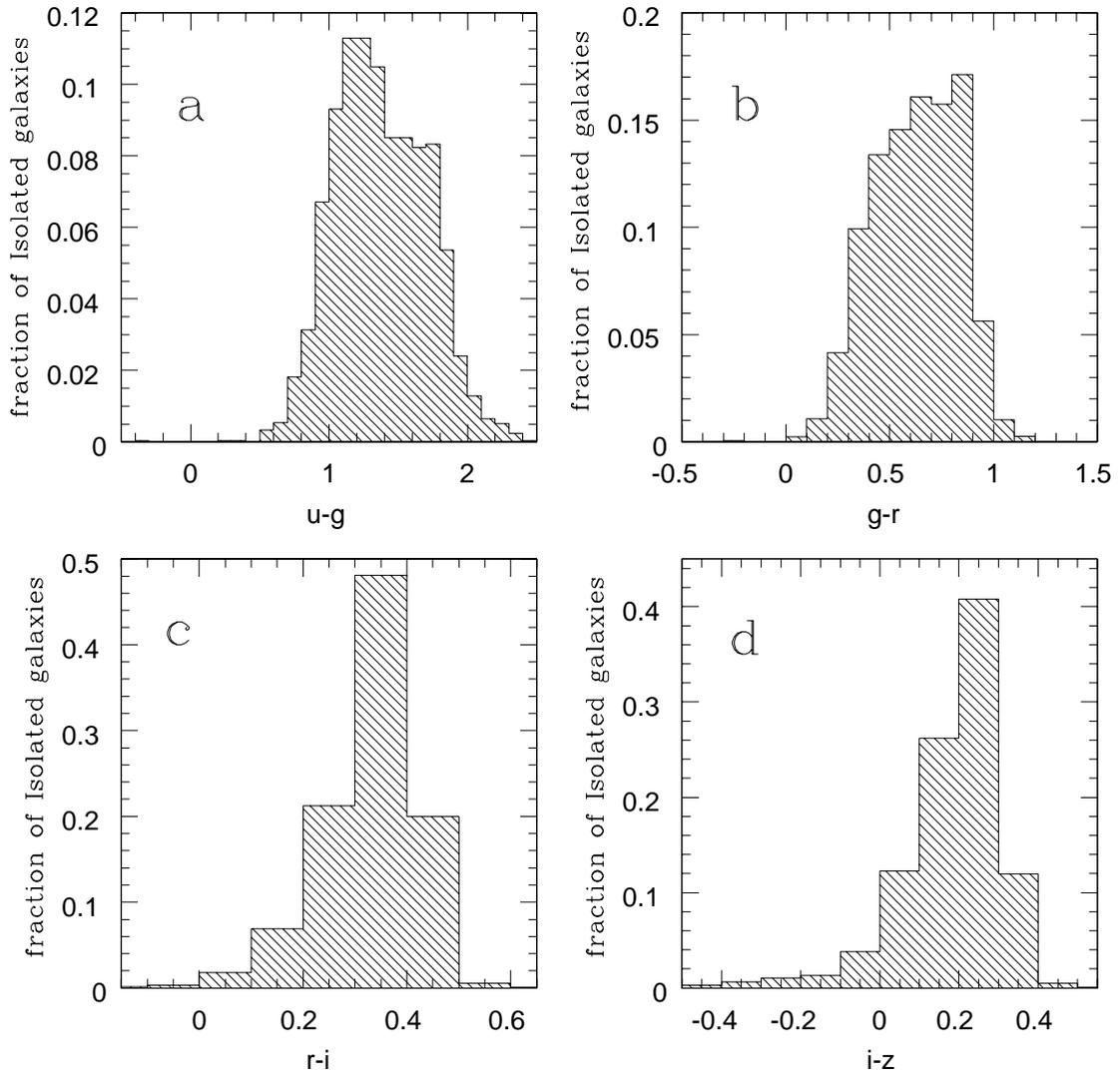,width=155mm,height=155mm,rwidth=150mm,clip=,angle=0,silent=}}
\caption{The distributions of the apparent colors of the SDSS DR1 isolated galaxies sample for 
$u-g$, $g-r$, $r-i$, and $i-z$.  Colors are corrected for Galactic extinction.}\label{figlabelhisCol}
\end{figure*}
\begin{figure*}
\centering
\makebox[160mm]{\psfig{file=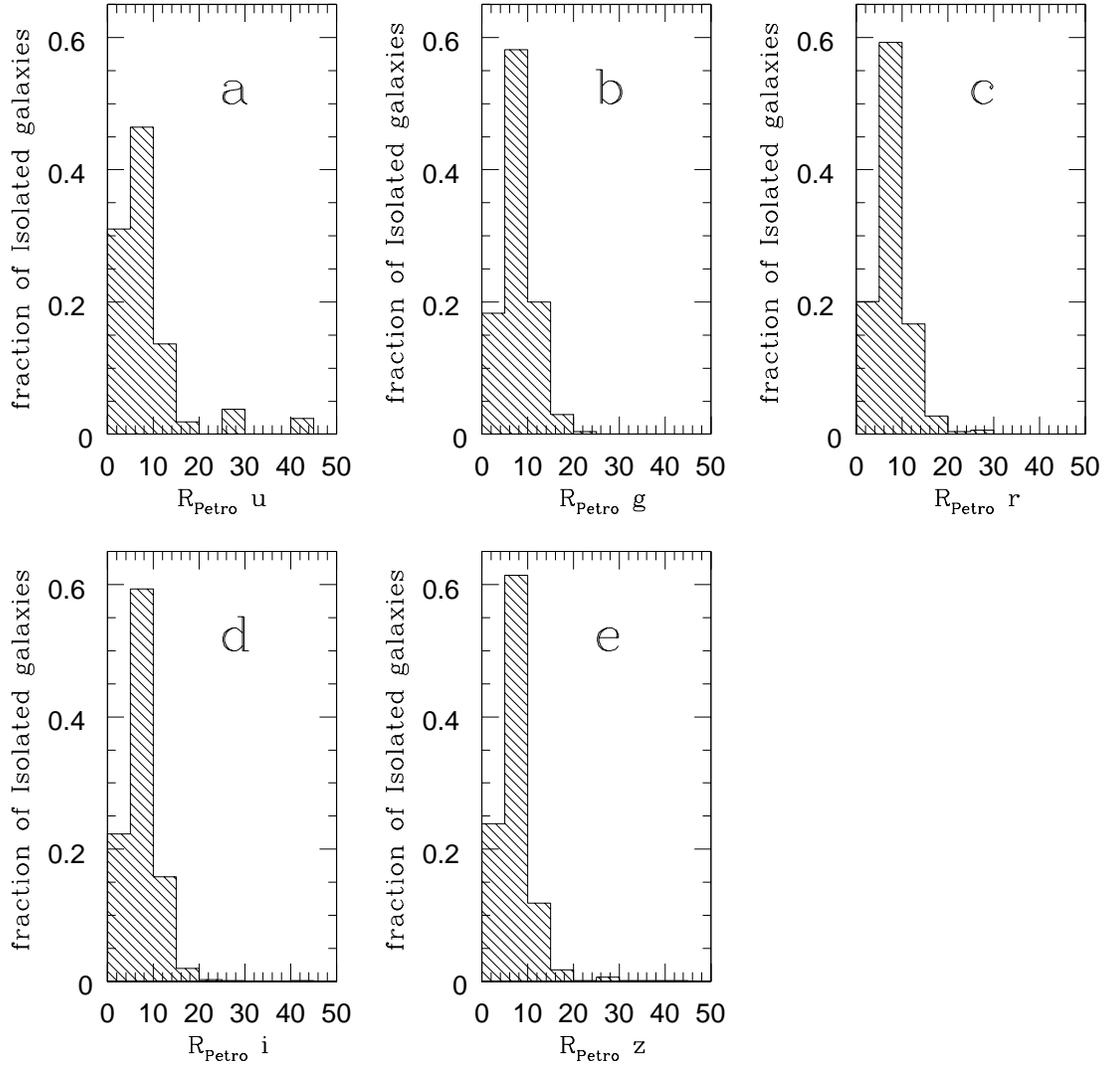,width=155mm,height=155mm,rwidth=150mm,clip=,angle=0,silent=}}
\caption{Distribution of the Petrosian radii of the SDSS DR1 isolated galaxies sample in the 5 SDSS filters.}
\label{figlabelhisradius}
\end{figure*}
\begin{figure*}
\centering
\makebox[50mm]{\psfig{file=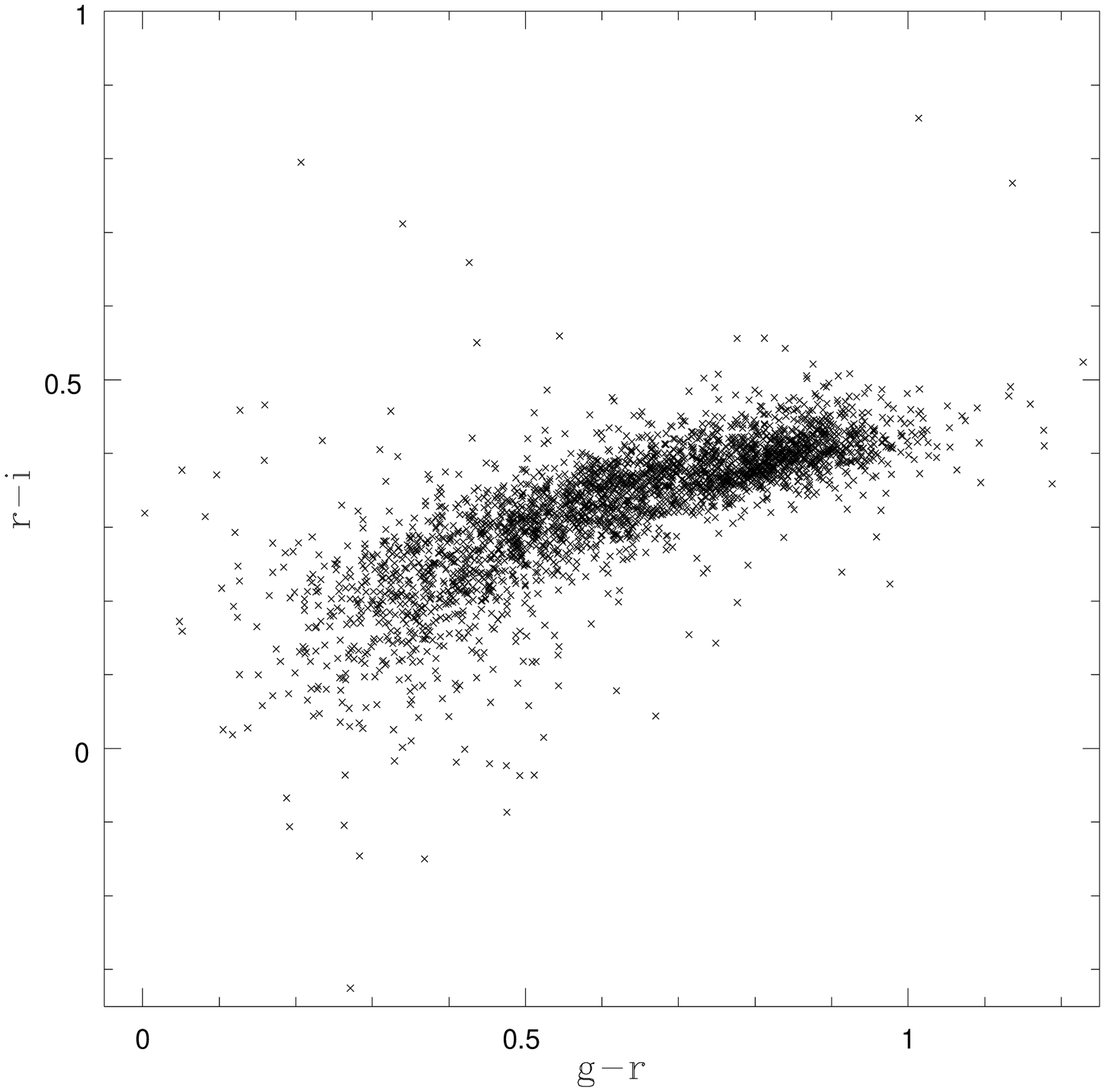,width=54mm,height=54mm,rwidth=45mm,clip=,angle=0,silent=}}
\makebox[50mm]{\psfig{file=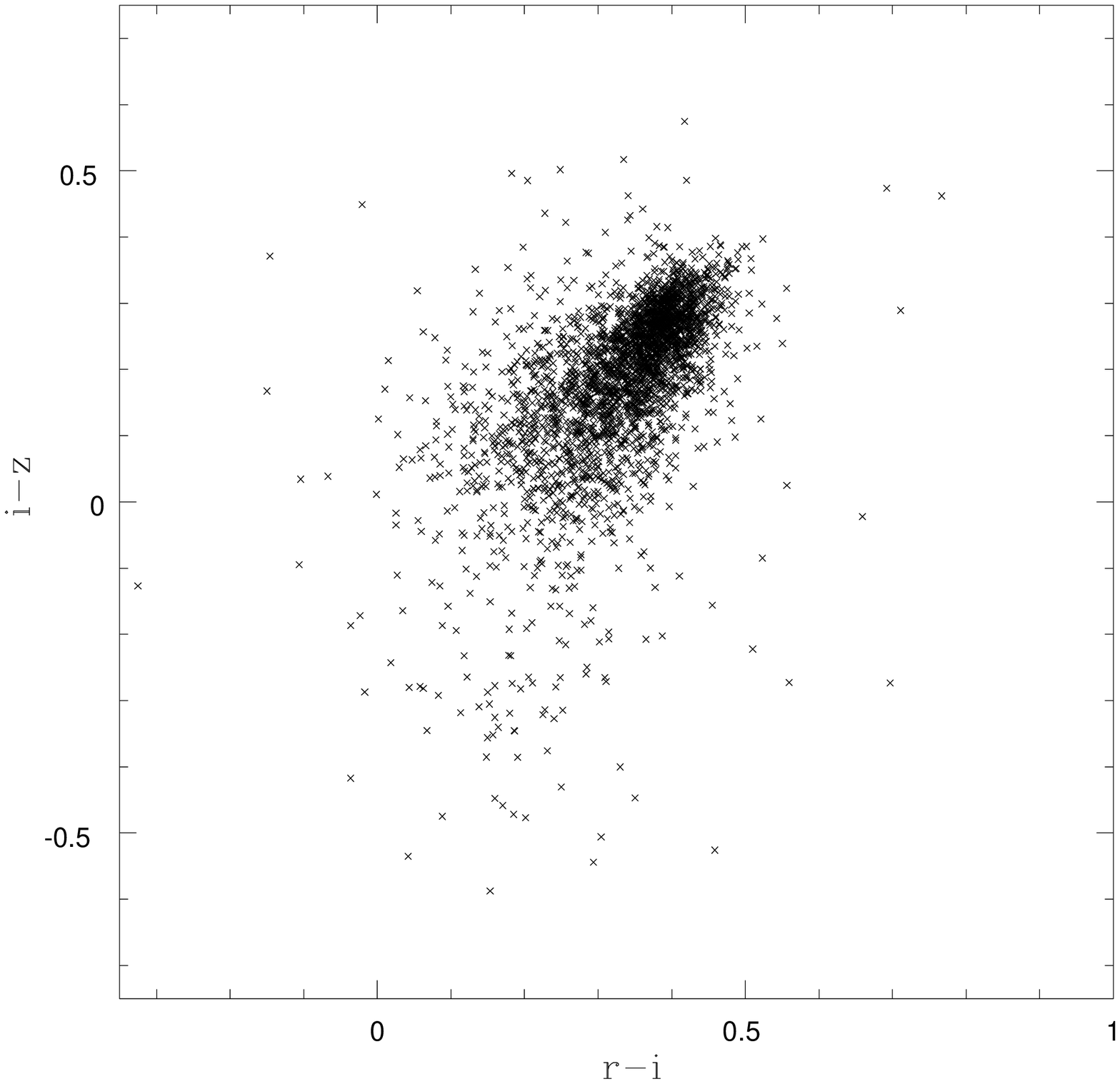,width=54mm,height=54mm,rwidth=45mm,clip=,angle=0,silent=}}
\makebox[50mm]{\psfig{file=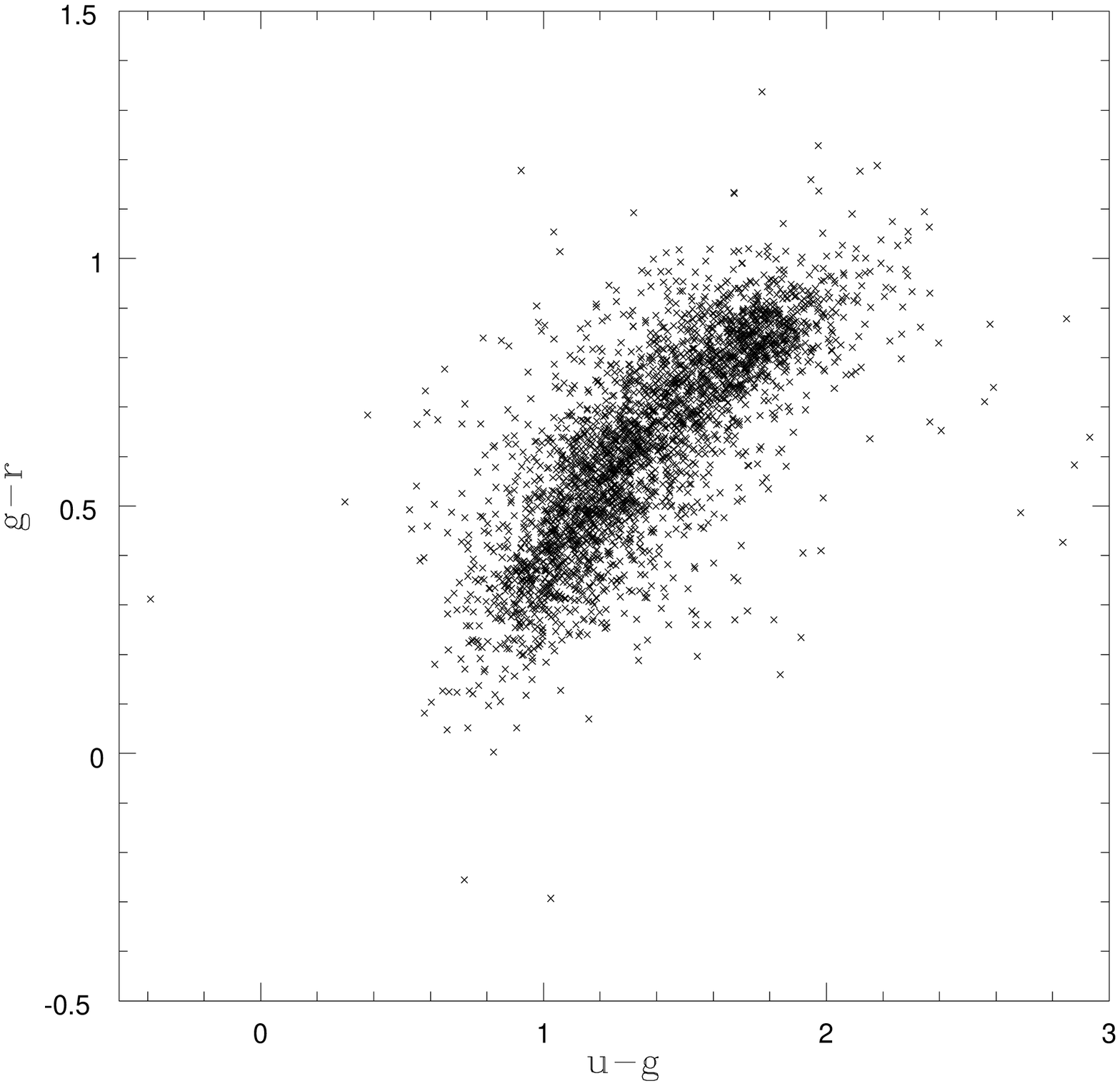,width=54mm,height=54mm,rwidth=45mm,clip=,angle=0,silent=}}
\caption{The (apparent) color-color diagrams for the SDSS DR1 isolated galaxies sample.}
\label{figlabelcol}
\end{figure*}
\begin{figure*}
\centering
\caption{Redshift-RA wedge plot of the sample of all SDSS DR1 galaxies with redshift (black dots) and of the sample 
of isolated galaxies that have redshift (red dots).}
\label{figlabelpolar}
\end{figure*}
\begin{figure*}
\caption{The k-corrected absolute $g-r$ colors for those isolated galaxies with a spectroscopic redshift as a function of redshift.  The colors have also been corrected for Galactic reddening from the Milky Way.}
\label{figlabelzMgMr}
\end{figure*}
\begin{figure*}
\centering
\caption{Absolute color-magnitude diagrams for the $g$ band for those isolated galaxies with a redshift. 
Colors and $g$ magnitudes have been k-corrected and reddening corrected.}
\label{figlabelMg}
\end{figure*}
\begin{figure*}
\centering
\caption{Absolute color-magnitude diagrams for the $r$ band for those isolated galaxies with a redshift. 
Colors and $r$ magnitudes have been k-corrected and reddening corrected.}
\label{figlabelMr}
\end{figure*}
\begin{figure*}
\centering
\caption{Absolute color-color diagrams for those isolated galaxies with a redshift. 
Colors have been k-corrected and reddening corrected.}
\label{figlabelMall}
\end{figure*}
\begin{figure*}
\centering
\makebox[160mm]{\psfig{file=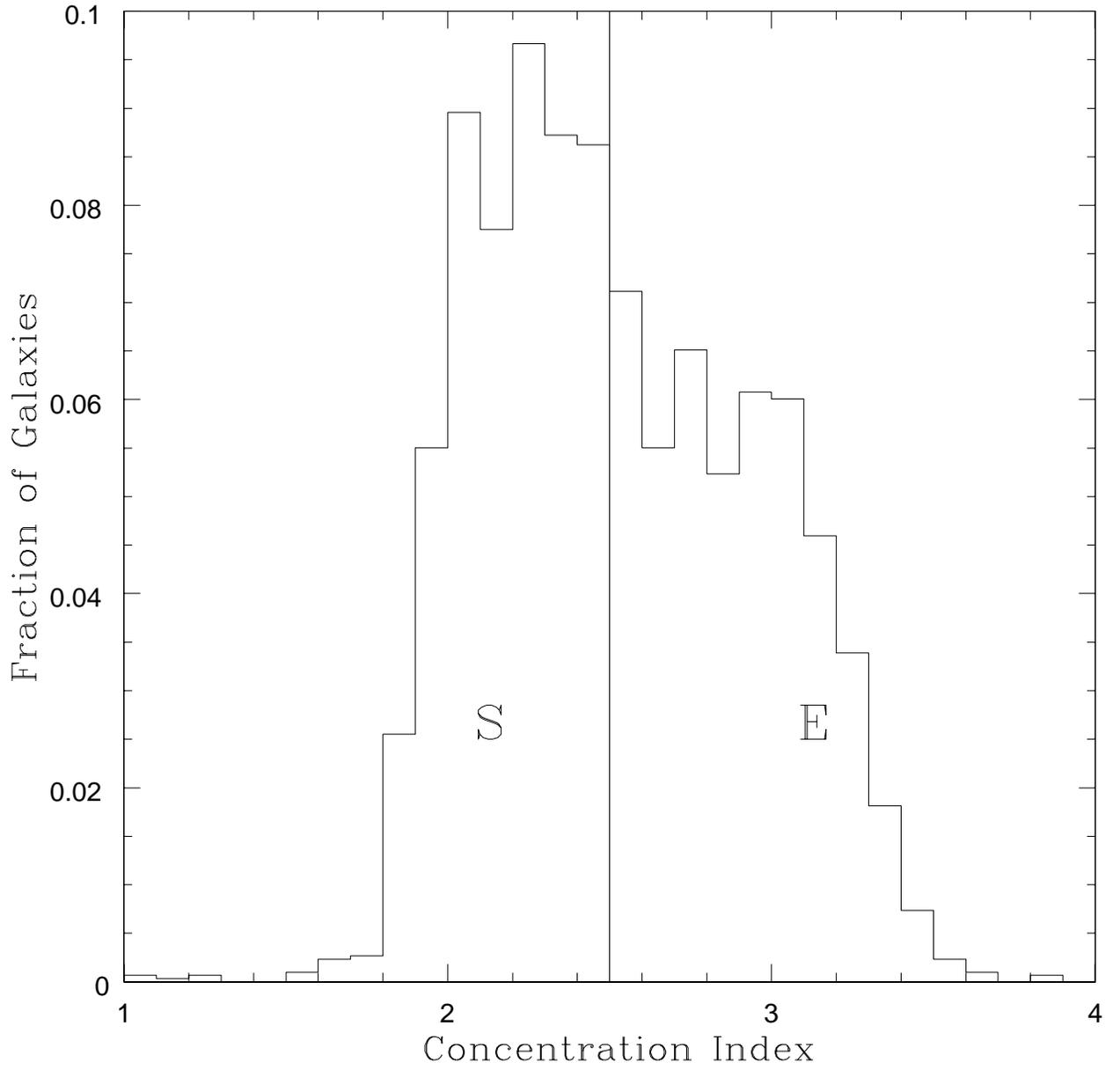,width=175mm,height=175mm,rwidth=150mm,clip=,angle=0,silent=}}
\caption{Distribution of concentration index (CI) for the isolated galaxies sample.}
\label{figlabeCi}
\end{figure*}
\begin{figure*}
\centering
\makebox[120mm]{\psfig{file=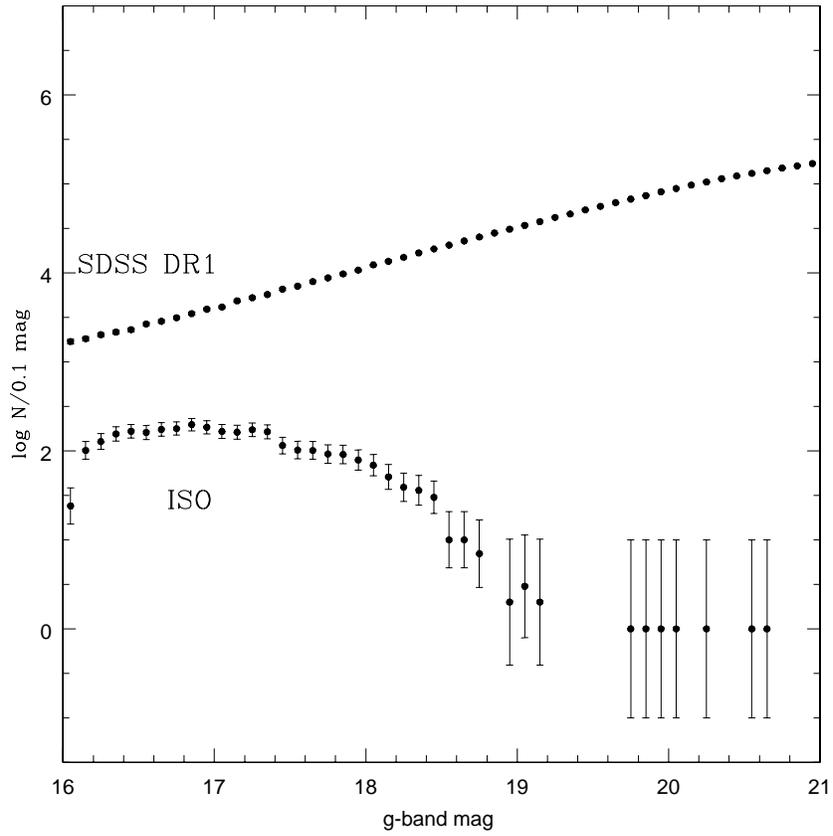,width=120mm,height=120mm,clip=,angle=0,silent=}}
\caption{The $g$ magnitude distribution in  0.1  magnitude bins of all 
SDSS DR1 galaxies and  of  the SDSS DR1  isolated galaxies.}
\label{figcomplet}
\end{figure*}  
\begin{figure*}
\centering
\makebox[120mm]{\psfig{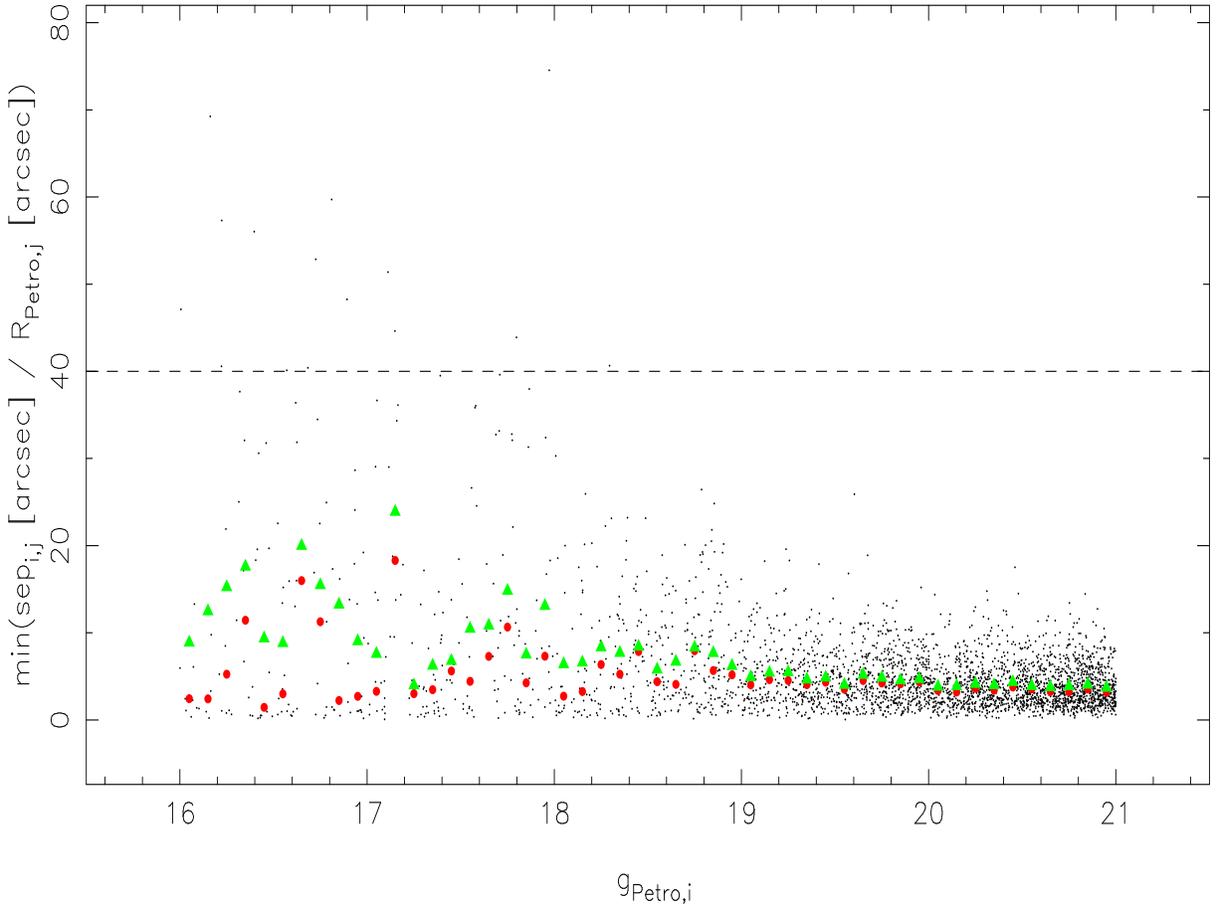}}
\caption{The  minimum value of  the scaled  separation $x_{i,j}/R_{j}$
(see eq.~3) vs.  $g$-band Petrosian magitude for each of 3354 galaxies
in a subset of the  SDSS DR1.  (For each  of these 3354 galaxies, only
neighbors within 3.0~mag  are considered; see  eq.~4.)  Black dots are
the individual values for $x_{i,j}/R_{j}$ vs. $g$; the green triangles
and  the red circles  are the mean and median  values in 0.1 mag bins,
respectively.  The dashed  horizontal   line at $x_{i,j}/R_{j} =   40$
indicates the  isolation   criterion of eq.   3;   any  galaxies whose
$x_{i,j}/R_{j}$  are plotted  above  this   line would be   considered
isolated.
\label{figCriterionTest}}
\end{figure*}  
\begin{figure*}
\centering
\makebox[160mm]{\psfig{file=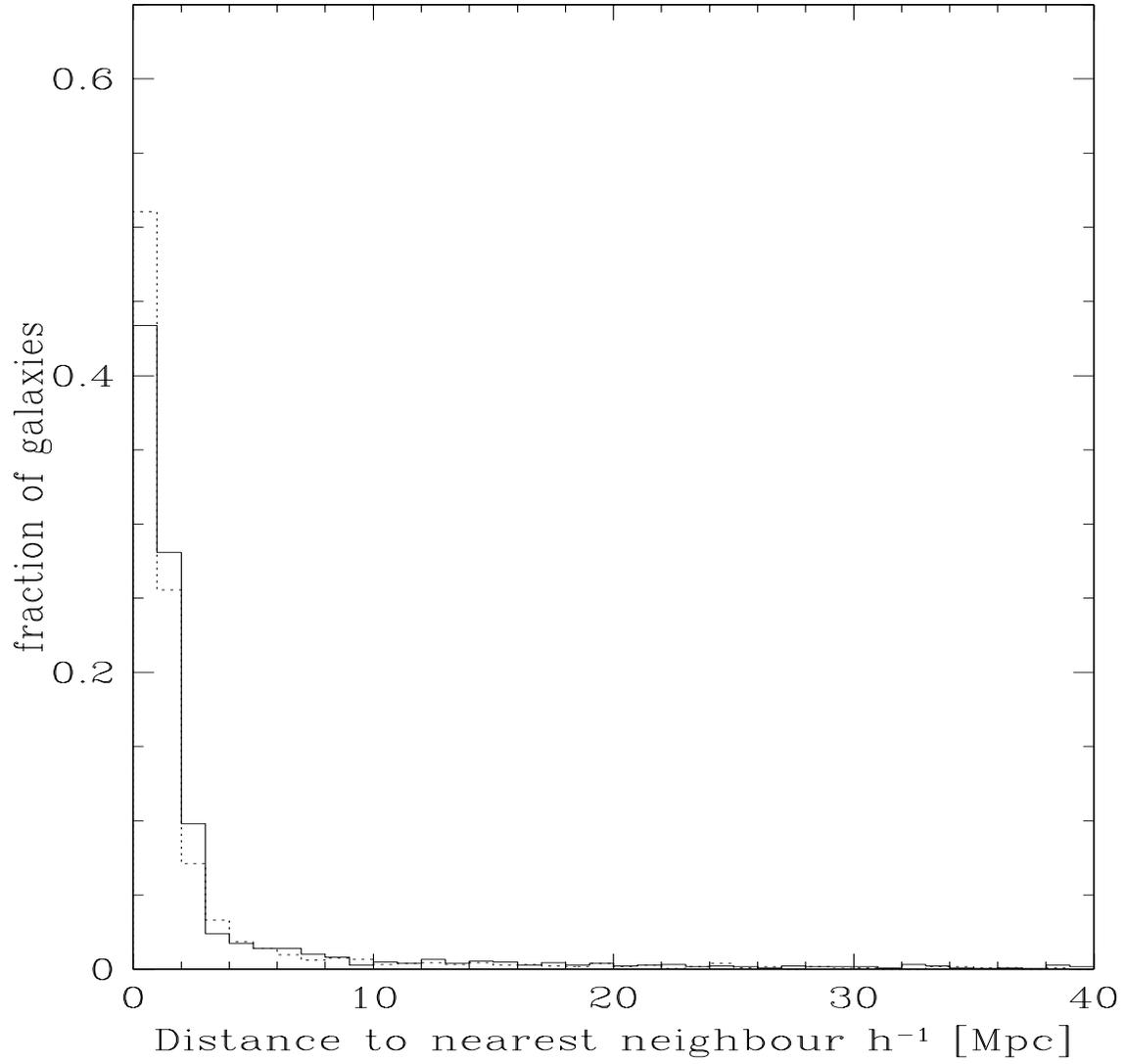,width=155mm,height=155mm,rwidth=150mm,clip=,angle=0,silent=}}
\caption{Distribution of nearest neighbor distances for the isolated galaxies sample.}
\label{figlabelneighbor}
\end{figure*}
\begin{figure*}
\centering
\makebox[160mm]{\psfig{file=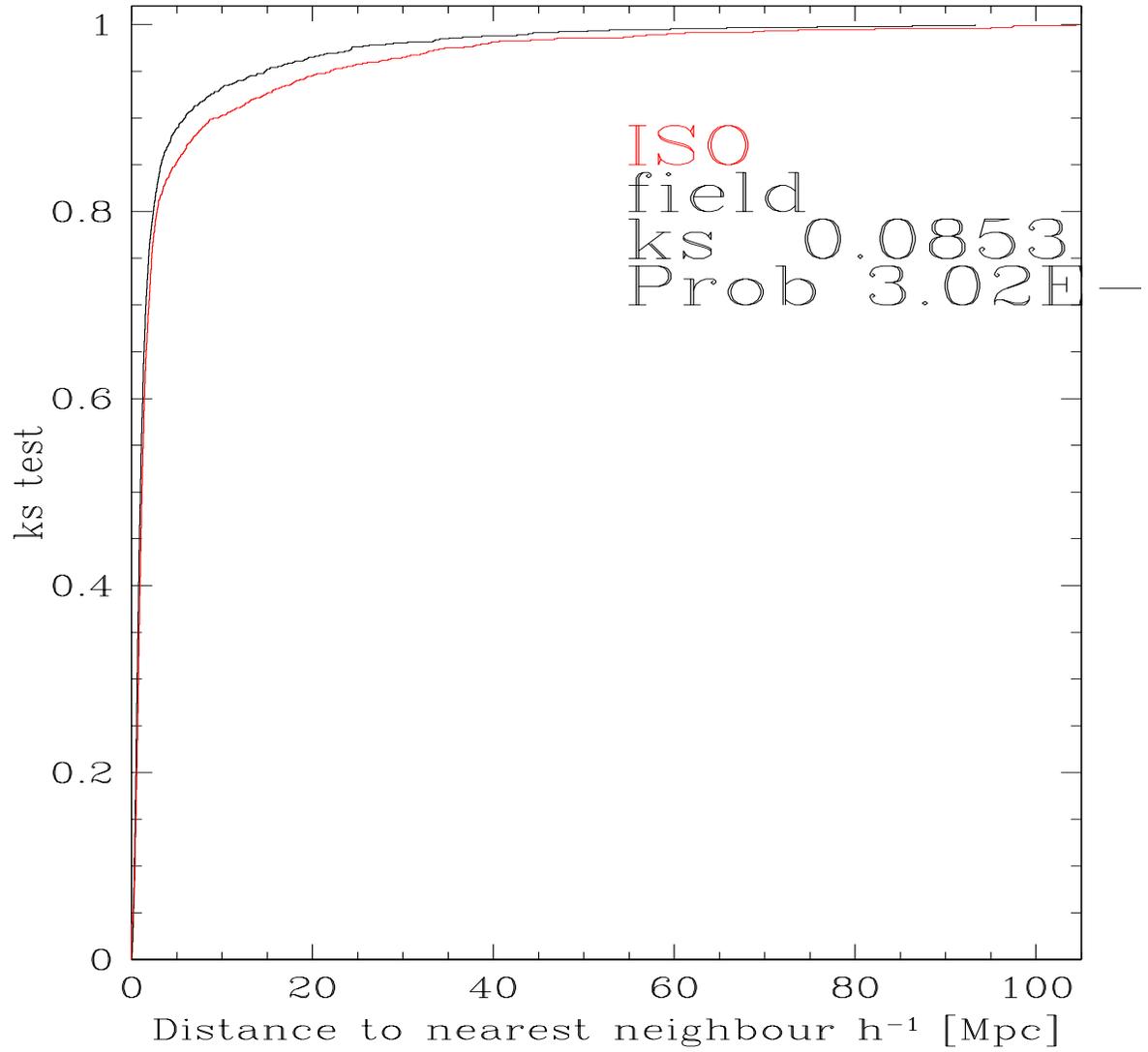,width=155mm,height=155mm,rwidth=150mm,clip=,angle=0,silent=}}
\caption{KS test comparing the distributions of nearest neighbor distances for the isolated galaxy sample
and for the field galaxy sample.}
\label{figlabelneighborKS}
\end{figure*}
%

\begin{thebibliography}{ }

\bibitem[Aars(2002)]{Aars02}
	Aars, C. E. 2002, Ph.D. Thesis

\bibitem[Aars(2003)]{Aars03}
	Aars, C. E.,  2003, American Astronomical Society Meeting, 203, \#65.01

\bibitem[Abazajian et al(2003)]{Abazajian03}
	Abazajian  K.,  et al.  2003, AJ, 126, 2081 

\bibitem[Adams, Jensen \& Stocke(1980)]{Adams80} 
	Adams M., Jensen E., Stocke J. 1980, AJ 85, 1010

\bibitem[Aikio \&  Maehoenen(1998)]{AikioMaehoenen98} 
	Aikio J.,  Maehoenen, P., 1998, ApJ 497, 534 

\bibitem[Bertin \& Arnouts(1996)]{BertinArnouts96}
	Bertin  E.,  Arnouts S.  1996, A\&AS 117, 393

\bibitem[Blanton et al(2003)]{Blanton03} 
	Blanton M. R., et al. 2003, AJ, 125, 2276

\bibitem[Eisenstein et al(2001)]{Eisenstein01}
	Eisenstein D. J., et al. 2001, AJ, 122, 2267

\bibitem[Frisch et al(1995)]{Frisch95}
	Frisch P., Einasto J., Einasto M.,  Freudling W., 
	Fricke K.J., Gramann M., Saar V., Toomet O., 
	1995, A\&A 296, 611

\bibitem[Fukugita et al(1996)]{Fukugita96}
	Fukugita, M., et al.  1996, AJ 111, 1748

\bibitem[Gunn et al(1998)]{Gunn98}
	Gunn, J. E., et al.  1998, AJ 116, 3040

\bibitem[Haynes \& Giovanelli(1980)]{Haynes80}
	Haynes, M. P.\& Giovanelli, R. 1980, \apjl, 240, L87 

\bibitem[Haynes, Giovanelli \& Chincarini(1984)]{Haynes84}
	Haynes, M. P., Giovanelli, R.,  Chincarini, G. L. 1984, 
	\araa, 22, 445 

\bibitem[Hogg et al(2001)]{Hogg01}
	Hogg, D. W., Finkbeiner, D. P., Schlegel, D.  J., Gunn, J. E.,
	2001, AJ, 122, 2129

\bibitem[Huchra \& Thuan(1977)]{HuchraThuan77}
	Huchra, J., Thuan, T. X.  1977, ApJ 216, 694  

\bibitem[Huchra et al.(2000)]{Huchra00}
	Huchra, J., et al. 2000, {\em CfA Redshift Catalog}

\bibitem[Karachentseva(1973)]{Karachentseva73}
	Karachentseva, V. E. 1973, 
	Soobshcheniya Spetsial'noj Astrofizicheskoj Observatorii, 8, 3 

\bibitem[Karachentseva(1980)]{Karachentseva80}
	Karachentseva, V. E. 1980, \azh, 57, 1153 
	(Soviet Astronomy, 24, 665)

\bibitem[Koopmann \& Kenney(1998)]{Koopmann98}
	Koopmann, R. A., Kenney, J. D. P. 1998, \apjl, 497, L75 

\bibitem[Leon \& Verdes-Montenegro(2003)]{Leon03} 
        Leon, S., Verdes-Montenegro, L. 2003, \aap, 411, 391 

\bibitem[Lindner et al(1995)]{Lindner95}
	Lindner U., Einasto J., Einasto M., Freudling  W., 
	Fricke K., Tago E., 1995, A\&A 301, 329 

\bibitem[Lisenfeld et al.(2003)]{2003sftt.conf..249L} 
	Lisenfeld,    U.,  Verdes-Montenegro,    L.,  Espada,      D.,
	Garc{\'{\i}}a, E., \& Leon, S.\ 2003, ASP Conf.~Ser.~297: Star
	Formation Through Time, 249

\bibitem[Lupton et al(2002)]{Lupton02}
	Lupton, R. H., Ivezic, Z., Gunn, J. E., Knapp, G., Strauss, M. A., 
	\& Yasuda, N. 2002, \procspie, 4836, 350

\bibitem[Pen(1999)]{Pen99}
	Pen, U.-L. 1999, \apjs, 120, 49


\bibitem[Pier et al(2003)]{Pier03} 
	Pier, J.  R., et al.  2003, AJ, 125,  1559 

\bibitem[Pisano et al.(2002)]{Pisano02}
	Pisano, D.,  Wilcots, E. M.,  \& Liu, C. T., 2002,  \apjs, 142, 161

\bibitem[Pisano \& Wilcots(2003)]{Pisano03}
	Pisano, D.,  \& Wilcots, E. M. 2003,  \apj, 584, 228

\bibitem[Prada et al.(2003)]{Prada03} 
	Prada, F.~et al.\ 2003,\apj, 598, 260 

\bibitem[Press et al.(1992)]{Press92} 
	Press,  W.  H.,   Teukolsky, S.   A., Vetterling,   W.  T., \&
	Flannery, B. P. 1992, Numerical Recipes in Fortran: The Art of
	Scientific  Computing,  2nd Edition,  (Cambridge:    Cambridge
	University Press)

\bibitem[Richards et al(2002)]{Richards02}
	Richards, G. et al. 2002, AJ, 123, 2945

\bibitem[Sauty et al.(2003)]{Sauty03}
	Sauty, S. et al. 2003, \aap, 411, 381

\bibitem[Schlegel, Finkbeiner \& Davis(1998)]{Schlegel98}
	Schlegel, D. J., Finkbeiner, D. P. \& Davis, M.
	1998, \apj, 500, 525

\bibitem[Shimasaku et al(2001)]{Shimasaku01}
	Shimasaku, K.,   et al.\ (2001), AJ, 122 1238

\bibitem[Smith et al.(2002)]{Smith02}
	Smith, J. A., et al. 2002, 123, 2121

\bibitem[Stavrev(1990)]{Stavrev90}
	Stavrev K.Y., 1990, Publ. Astron. Dept. Eötvös Univ. 10, 115 

\bibitem[Stocke et al.(2004)]{Stocke04}
	Stocke, J. T., Keeney, B. A., Lewis, A. D., Epps, H. W., \& 
	Schild, R. E.,  2004, \aj, 127, 1336

\bibitem[Stoughton et al(2002)]{Stoughton02}
	Stoughton, S. et al.  2002, AJ 123, 485 

\bibitem[Strauss et al(2002)]{Strauss02}
	Strauss, M. A., et al. 2002, AJ, 124, 1810

\bibitem[Turner  \& Gott(1975)]{TurnerGott75}
	Turner, E., Gott, J.R., 1975, ApJ, 197L, 89

\bibitem[Tully(1988)]{Tully88}
	Tully, R. B. 1988, ``Nearby Galaxies Catalog,'' 
	Cambridge and New York, Cambridge University Press

\bibitem[Varela et al.(2004)]{Varela04}
	Varela, J., Moles, M.,  M\'arquez, I., Galleta,  G., Masegosa,
	J., \& Bettoni, D.  2004,  A\&A, 420, 873

\bibitem[Yasuda et al(2001)]{Yasuda01}
	Yasuda, N., et al.\ 2001, AJ, 122 1104

\bibitem[York et al(2000)]{York00}
 	York, D. G., et al.  2000, AJ 120, 1579

\bibitem[Zwicky et al.(1968)]{Zwicky68}
	Zwicky, F., et al.  1968, Catalogue of Galaxies \& Clusters of Galaxies
\end{thebibliography}
\end{document}